# Transport properties in magnetized compact stars

**Toshitaka Tatsumi [1,*], Hiroaki Abuki [2]**

[1] Institute of Education, Osaka Sangyo University, 3-1-1 Nakagaito, Daito, Osaka 574-8530, Japan; tatsumitoshitaka@gmail.com
[2] Aichi University of Education, 1 Hirosawa 1, Kariya 448-8542, Japan ; abuki@auecc.aichi-edu.ac.jp
\* Correspondence: tatsumitoshitaka@gmail.com.

**Abstract:** Transport properties of dense QCD matter is discussed. Using the Kubo formula for conductivity, we discuss some topological aspects of quark matter during chiral transition. The close relation to Weyl semimetal is pointed out and anomalous Hall effect is demonstrated to be possible there. In particular, it is shown that the spectral asymmetry of the quasi-particles plays an important role for the Hall conductivity in the magnetic field.

## 1. Introduction

Nowadays many efforts have been made theoretically or experimentally to explore the QCD phase diagram in the temperature-density plane [1], where various phase transitions have been proposed. Among them deconfinement transition and chiral transition are fundamental subjects for QCD. Phenomenologically these transitions may affect the evolution of the universe or physics of compact stars. We, hereafter, consider the chiral transition mainly at low temperature.

Many studies have shown that chiral transition from the spontaneously symmetry broken (SSB) phase to the quark-gluon plasma (QGP) phase should be present at some density and temperature, using the effective models of QCD, while at the moment it has not yet been directly proven from the first principle computations with lattice gauge theory due to the sign problem [2]. Recently, possible appearance of the inhomogeneous chiral phase (iCP) has been suggested near the chiral transition, and extensively studied in various situations, including in the presence of magnetic field [3-7].

Here we reveal a new aspect of iCP in relation to Weyl semimetal (WSM) in condensed-matter physics [8]. We focus on the transport properties in iCP to reveal some topological features inherent in iCP. First, we shall see that energy spectrum of quasi-particles in the iCP phase is the same with WSM, and anomalous Hall effect (AHE) [9,10] also works in iCP. In this case AHE is brought about by the magnetic monopole in the momentum space. Next, we consider the Hall conductivity in the presence of magnetic field. This subject is important to understand the transport properties inside compact stars such as thermal evolution of magnetars [11]. In condensed-matter physics this subject has been also discussed to elucidate the transport properties of topological materials [12]. We shall see that there also appears anomalous Hall conductivity as the magnetic-field independent term due to the spectral asymmetry. So the competition between AHE and the usual Hall effect is interesting as the strength of magnetic field increases.

## 2. Inhomogeneous chiral phase (iCP) in dense QCD and Weyl semimetal

*2.1. Dual chiral density wave (DCDW)*

iCP is characterized by the generalized order parameter,

$$\chi(\mathbf{r}) \equiv \langle \bar{\psi}\psi \rangle + i \langle \bar{\psi} i \gamma_5 \tau_3 \psi \rangle = \Delta(\mathbf{r}) \exp(i\theta(\mathbf{r})) \quad (1)$$

for the symmetry breaking of $SU(2)_L \times SU(2)_R$. The order parameters $\Delta, \theta$ are now spatially dependent, and various types of $\chi(\mathbf{r})$ can be chosen, depending on the situation [5].

The *dual chiral density wave* (DCDW) is a kind of density wave and the DCDW phase is specified by the scalar and pseudoscalar condensates with spatial modulation [3],

$$\langle \bar{\psi}\psi \rangle = \Delta \cos(\mathbf{q} \cdot \mathbf{r}),$$
$$\langle \bar{\psi} i \gamma_5 \tau_3 \psi \rangle = \Delta \sin(\mathbf{q} \cdot \mathbf{r}) \quad (2)$$

The order parameters are the amplitude $\Delta$ and the wave-vector **q**. We, hereafter, focus on this type, because the phase degree of freedom is indispensable for manifestation of topological effects in iCP. Moreover, it may be most favorite configuration in the presence of the magnetic field [5-7].

Taking the Nambu-Jona-Lasinio (NJL) model as an effective model of QCD at low energy scale.

$$L_{\text{NJL}} = \psi(i\slashed{\partial} - m_c)\psi + G\left[(\bar{\psi}\psi)^2 + (\bar{\psi} i \gamma_5 \boldsymbol{\tau} \psi)^2\right] \quad (3)$$

we, hereafter, consider the flavor-symmetric *u* and *d* quark matter in the chiral limit, $m_c = 0$. Then, the effective Lagrangian can read

$$L_{MF} = \bar{\psi}\left(i\slashed{\partial} - \frac{1+\gamma_5\tau_3}{2}M(z) - \frac{1-\gamma_5\tau_3}{2}M^*(z)\right)\psi - \frac{|M(z)|^2}{4G},$$
$$M(z) = -2G\Delta e^{iqz} (\equiv M e^{iqz}) \quad (4)$$

under the mean-field approximation. One may rewrite it in a simple form,

$$L_{MF} = \bar{\psi}_W\left[i\slashed{\partial} - M - 1/2\gamma_5\tau_3 \slashed{q}\right]\psi_W - G\Delta^2 \quad (5)$$

with $M = -2G\Delta$ and the space-like vector $q^\mu = (0, \mathbf{q})$ by the use of the Weinberg transformation, $\Psi_W = \exp[i\gamma_5\tau_3\mathbf{q}\cdot\mathbf{r}/2]$. Thus we can see that the amplitude of DCDW provides the dynamical mass for the newly defined quarks (quasi-particles) described by $\Psi_W$, while the wave-vector plays a role of the mean-field for them. The effective Hamiltonian operator in the momentum space then renders

$$H_{MF}(\mathbf{p}) = \begin{pmatrix} M + \tau_3/2\boldsymbol{\sigma}\cdot\mathbf{q} & \boldsymbol{\sigma}\cdot\mathbf{p} \\ \boldsymbol{\sigma}\cdot\mathbf{p} & -M + \tau_3/2\boldsymbol{\sigma}\cdot\mathbf{q} \end{pmatrix}. \quad (6)$$

The single-particle energy can be easily extracted,

$$E_{\varepsilon=\pm 1, s=\pm 1}(p) = \varepsilon\sqrt{E_p^2 + |\mathbf{q}|^2/4 + s\sqrt{(\mathbf{p}\cdot\mathbf{q})^2 + M^2|\mathbf{q}|^2}}, \quad (7)$$

with $E_p = \sqrt{p^2 + M^2}$ for each flavor, where $\varepsilon$ and $s$ denote the particle-antiparticle and spin degrees of freedom, respectively. Accordingly, this form suggests anisotropy of the Fermi sea in the momentum space: it deforms in the axial-symmetric way around the direction of **q**.

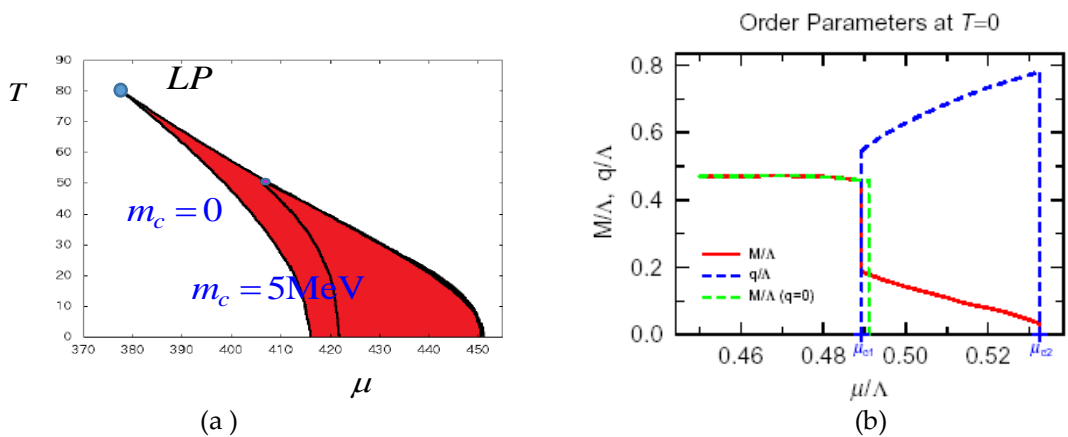

(a)  (b)

**Figure 1.** (a) QCD phase diagram including the DCDW phase in the density-temperature plane, taken from ref.[13]. The triple point *LP*, is called the Lifshitz point, where the SSB phase, the QGP phase and iCP coexsist; (b) Density dependence of the order parameters in the unit of the cut-off parameter $\Lambda$, taken from ref.[3]. The dynamical mass in the usual chiral transition is also depicted by $M/\Lambda$ ($q = 0$).

Thermodynamic potential can be easily evaluated for given temperature $T$ and baryon-number chemical potential $\mu$, using the single-particle energy (7). Then, the values of the order parameters can be easily extracted: we can see that the DCDW phase appears around several times the normal nuclear density at $T = 0$ (see Fig. 1 (a)), where the usual chiral transition has been expected [3]. It should be interesting to see that the value of the wave-vector becomes rather high, $q/2 = \mathcal{O}(\mu) > M$ (Fig.1 (b)), which is reminiscent of the nesting effect of the Fermi surface. Actually it has been shown that the DCDW phase is always favored by the nesting effect at 1+1 dimension [13,14]. We also anticipate that the appearance of the phase with spatially modulating order should be rather universal phenomenon during the phase transitions both described by the uniform order parameter such as FFLO state in the context of superconductivity [15,16].

*2.2. Similarity with Weyl semimetal (WSM)*

The energy spectrum for $s = 1$ has a gap between positive and negative energies in momentum space. On the other hand, the energy spectrum for $s = -1$ exhibits an interesting behavior, depending on the values of $q$ and $M$; for $q/2 < M$ there is an energy gap, while there is no energy gap at the points, $\mathbf{p} = (p_x, p_y, \pm K_0)$, with $K_0 = \sqrt{(q/2)^2 - M^2}$ for the opposite case, $q/2 > M$. These points are called Weyl points in condensed-matter physics and the quasi-particle excitations can be expressed by the use of the $2 \times 2$ Weyl Hamiltonian around these points. The latter case is actually realized in the DCDW phase. In Fig. 2 we sketch the energy surface for $s = -1$ and $q/2 > M$.

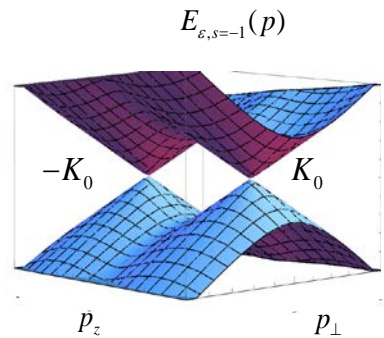

$E_{\varepsilon,s=-1}(p)$

**Figure 2.** Schematic view of the energy surface of the quasi-particles with $s = -1$ in the momentum space. The Weyl points are located at $\pm K_0$ on the $p_z$ axis, around which the excitations of the quasi-particles are described by the Weyl Hamiltonian.

Here we'd like point out a similarity to Weyl semimetal (WSM). WSM is one of the topological materials and can be modelled by the Dirac type Hamiltonian [7],

$$H_{\text{WSM}}(\mathbf{k}) = \begin{pmatrix} m + b\sigma_z & \mathbf{k} \cdot \boldsymbol{\sigma} \\ \mathbf{k} \cdot \boldsymbol{\sigma} & -m + b\sigma_z \end{pmatrix}, \quad (8)$$

where $\mathbf{k}$ denotes the Bloch momentum, and $m$ is the spin-orbit coupling strength. The parameter $b$ in the CPT violating term controls the spin-splitting due to magnetic impurities doped in the materials. Thus one may immediately notice that Hamiltonian (8) has the same structure with (6) for the DCDW case; $b$ corresponds to $q/2$. Some remarks are in order here: first, the parameter $m$ can take $m = 0$ without any inconsistency, in highly contrast to the DCDW phase, where $M = 0$ implies no phase transition. Secondly, only the negative-energy states (valence band) are occupied in WSM, while some Fermi sea (conduction band) is formed in the DCDW phase. Thirdly, there is a boundary in the WSM.

This is a new encounter of nuclear physics and condensed-matter physics in the context of compact stars. One may extract some hints or suggestions to transport properties in dense quark matter by way of studies of topological materials such as WSM.

## 3. Anomalous Hall effect (AHE) in iCP

### 3.1. Hall conductivity

Once we notice similarities with WSM, what can we learn about the DCDW phase? We, hereafter, examine the possible anomalous Hall effect (AHE) in the DCDW phase. When magnetic field is applied, off-diagonal resistivity $\rho_{xy} \propto [\sigma^{-1}]_{xy}$ is generally expressed as $\rho_{xy} = R_0 B_z + R_s M_z$ with magnetization **M**. The first term is a usual one representing the Hall effect, while the second term implies that the Hall effect occurs without any magnetic field. Such effect is called anomalous Hall effect (AHE) [8,9]. AHE provides one of most firm experimental evidences of WSM [8]. It may affect the transport properties through modification of the Maxwell equations [18,19]. Its phenomenological implications should be also interesting for compact stars. In the following we'd like to demonstrate AHE in the DCDW phase, as in WSM. We shall see that the intrinsic contribution to AHE can be regarded as an "unquantized" version of the quantum Hall effect.

We can derive the Hall coefficient by way of the Kubo formula, considering a linear response to a tiny electric field [20]. For translational invariant systems, the Hall conductivity renders

$$\sigma_{xy} = e^2 \int \frac{d^3k}{(2\pi)^3} B_z(\mathbf{k}) f(E_k), \tag{9}$$

Where $\mathbf{B}(\mathbf{k})$ is the Berry curvature in the momentum space, defined by the Berry connection $\mathbf{A}(\mathbf{k})$, $\mathbf{B}(\mathbf{k}) = \nabla_k \times \mathbf{A}(\mathbf{k})$. It is to be noted that the factor $e^2$ comes from the cancellation due to the different directions of the wave vector for $u$ and $d$ quarks. Once the eigenfunctions $|u_k\rangle$ are known for any Hamiltonian $H(\mathbf{k})$, the Berry connection renders

$$A(\mathbf{k}) = -i\langle u_k | \nabla_k | u_k \rangle. \tag{10}$$

When Eq. (9) is further rewritten as

$$\sigma_{xy} = \int \frac{dk_z}{2\pi} \left[ e^2 \int \frac{d^2k}{(2\pi)^2} B_z(\mathbf{k}) f(E_k) \right] \tag{11}$$

we find that the formula in the parenthesis is the TKNN formula of 2D quantum Hall systems [21], where the Hall conductivity can be written as a momentum integral over the Brillouin zone,

$$\sigma_{xy} = e^2 \sum_n \int_{BZ} \frac{d^2k}{(2\pi)^2} \left[ \frac{\partial A_{n,y}}{\partial k_z} - \frac{\partial A_{n,z}}{\partial k_y} \right]$$

$$= \frac{e^2}{2\pi} \sum_n \nu_n. \tag{13}$$

The integer $\nu_n$ is a topological quantity called the first Chern number.

### 3.2. AHE in the DCDW phase

Since the eigenenergy and eigenfunction are already obtained, it is straightforward to evaluate the Berry curvature in the DCDW phase. Then its $z$-component renders

$$B_{s,z} = \frac{-1}{2E^3_{\varepsilon=+1,s}} \left( sE_0 + \frac{q}{2} \right), \tag{14}$$

with $E_0 = \sqrt{p_z^2 + M^2}$. For $s = -1$, there appear two Weyl points at $\mathbf{p} = (p_x, p_y, \pm K_0)$, and we can easily see that $\mathbf{B}_{s=-1}$ resembles the magnetic field produced by the Dirac monopoles sitting at the Weyl points; e.g. Eq.(14) takes the form, $B_{-1,z} \cong \frac{p'_z}{2|\mathbf{p}'|^3}$, $B_{-1,z} \cong \frac{p'_z}{2|\mathbf{p}'|^3}$ with $\mathbf{p}' = (p_x, p_y, p_z - K_0)$ around one Weyl point, $\mathbf{p} = (p_x, p_y, K_0)$.

Then, the contribution to the Hall conductivity from the Dirac sea can be evaluated as

$$\sigma_{xy}^{\text{Dirac}} = e^2 \sum_{s=\pm 1} \int \frac{d^3 p}{(2\pi)^3} \frac{1}{2 E_{\varepsilon=+1s}^3} (sE_0 + q/2). \tag{15}$$

The integral apparently diverges and depends on the cut-off scheme [22,23]. It is a subtle point; we must impose a physical condition to fix the finite term, which may be regarded as a kind of the renormalization condition. The relevant condition for the DCDW phase is $\sigma_{xy}^{\text{Dirac}} \to 0$ as $M \to 0$, because AHE should be vanished at the normal quark matter. This is achieved by using the gauge invariant regularizations such as the proper-time method or the heat-kernel method. Here we apply the proper-time regularization,

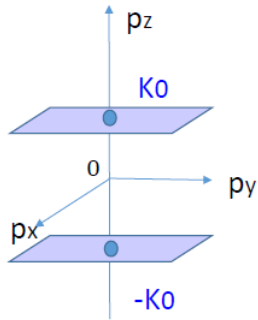

**Figure 3.** Integral along the $p_z$ axis. The function $sE_0 + q/2$ changes its sign at the Weyl points for $s = -1$, while it is constant for $= 1$.

$$\sigma_{xy}^{\text{Dirac}} = \lim_{\Lambda \to \infty} \frac{e^2}{2\Gamma(3/2)} \sum_{s=\pm 1} \int \frac{d^3 p}{(2\pi)^3} (sE_p + q/2) \int_{\Lambda^{-2}}^{\infty} d\tau \tau^{1/2} e^{-\tau E_s^2}$$

$$= \frac{e^2}{4\pi^2} \sum_{s=\pm 1} \int_0^{\infty} dp_z \text{sign}(sE_0 + q/2)$$

$$- \frac{e^2}{4\pi^{5/2}} \lim_{\Lambda \to \infty} \sum_{s=\pm 1} \int_0^{\infty} dp_z (sE_0 + q/2) \int_0^{\Lambda^{-2}} d\tau \tau^{-1/2} e^{-\tau(sE_0 + q/2)^2} \tag{16}$$

The second term is evaluated to give $-e^2 q/(2\pi)^2$, while the first term gives,

$$\frac{e^2}{4\pi^2} (2K_0). \tag{17}$$

One may find this is related to the topological quantity by rewriting it as

$$\int_{-K_0}^{K_0} \frac{dp_z}{2\pi} \left( \frac{e^2}{2\pi} \right) \tag{18}$$

the quantity $e^2/2\pi$ is nothing else but the Hall conductivity for 2D quantum Hall systems with the Chern number $\nu = 1$. Thus we can regard the DCDW phase as a stack of 2D quantum Hall systems to the 3rd direction [21]. Eventually we find

$$\sigma_{xy}^{\text{Dirac}} = \frac{e^2}{4\pi^2} (2K_0 - q), \tag{19}$$

where we can easily check $\sigma_{xy}^{\text{Dirac}} \to 0$ as $M \to 0$, as it should be. Accordingly, the anomalous Hall current can be represented in the form $\mathbf{j}_{\text{AHE}}^{\text{Dirac}} = \frac{e^2}{(2\pi)^2} (1 - 2K_0/|\mathbf{q}|) \mathbf{q} \times \mathbf{E}$. It is to be noted that this expression is somewhat different for the one for Weyl semimetal, where the contribution from the valence band reads $(\sigma_{xy})_{WSM} = e^2 K_0/2\pi^2$. This is because AHE is possible there even if $m = 0$.

## 4. Transport properties in the presence of the magnetic field

*4.1. Hall conductivity*

The effective Hamiltonian for quasi-particles in the DCDW phase is given as

$$H_{MF} = \boldsymbol{\alpha} \cdot (\mathbf{p} - Q\mathbf{A}) + \beta M - \mathbf{q} \cdot \boldsymbol{\alpha} \gamma_5 \tau_3 / 2, \quad (20)$$

with $Q = \text{diag}(e_u, e_d)$ ($e_u = 2e/3, e_d = -e/3$), in the presence of the magnetic field. Since it has been shown that $\mathbf{q} \parallel \mathbf{B}$ is the most favorable configuration [24], we set magnetic field along z-axis without loss of generality.
Then, the Kubo formula gives the formula of the Hall conductivity [25],

$$\sigma_{xy} = i \int_{-\infty}^{+\infty} f(\eta) A_{xy}(\eta) d\eta,$$

$$A_{xy}(\eta) = i \text{Tr}\left\{ Q^2 \left[ j_x \left( dG^+/d\eta \right) j_y \delta(\eta - H_{MF}) - j_x \delta(\eta - H_{MF}) j_y \left( dG^-/d\eta \right) \right] \right\}, \quad (21)$$

with $f(\eta)$ being the Fermi-Dirac distribution function, where the Green functions are defined by $G^\pm(\eta) = (\eta - H_{MF} \pm i\varepsilon)^{-1}$, and the current operator $j_a$ is given by $j_a = -i[r_a, H_{MF}] = \alpha_a$.

Streda has further rewritten it into the following form [26],

$$\sigma_{xy} = \sigma_{xy}^{I} + \sigma_{xy}^{II}, \quad (22)$$

where the first term corresponds to the classical Drude-Zener result,

$$\sigma_{xy}^{I} = \frac{1}{2} i \text{Tr}\left\{ Q^2 \left[ j_x G^+(\mu) j_y \delta(\mu - H) - j_x \delta(\mu - H) j_y G^-(\mu) \right] \right\}, \quad (23)$$

at $T = 0$, and the second term can be expressed by the use of the number of state $N^f(E)$ for each flavor to represent the quantum effect,

$$\sigma_{xy}^{II} = -\sigma_{yx}^{II} = \sum_f e_f \left. \frac{\partial N^f(E)}{\partial B} \right|_{E=\mu}. \quad (24)$$

The number of state below the energy $E$ is defined by

$$N^f(E) = \frac{1}{2} \int_{-\infty}^{E} \text{Tr}' \delta(\eta - H_{MF}) d\eta, \\ = N^f(0) + N_F^f(E) \quad (25)$$

where $\text{Tr}'$ means no summation in the flavor space. It consists of the contribution of the Dirac sea $N^f(0)$ and that of the Fermi sea $N_F^f(E)$. In the usual case the first term gives no contribution and only the second term gives rise to the Hall conductivity. On the other hand, anomalous Hall conductivity just comes from the first term, when it is non-vanishing. We'd like to demonstrate it.

*4.2 Spectral asymmetry and AHE*

The effective Hamiltonian can be easily diagonalized [24] to give

$$E_{n,s,\varepsilon}^f(p_z) = \varepsilon \sqrt{\left( s\sqrt{M^2 + p_z^2} + q/2 \right)^2 + 2|e_f|Bn}, \quad n = 1,2,3,... \\ E_{n=0,\varepsilon}(p_z) = \varepsilon \sqrt{M^2 + p_z^2} + q/2 \quad (26)$$

for each flavor $f$, where $\varepsilon = \pm 1$ denotes the particle-antiparticle states, and $s = \pm 1$ specifies the spin degree of freedom. Note that there is no spin degree of freedom for the lowest Landau level (LLL), $= 0$ : there occurs dimensional reduction for LLL and the eigenspinor is represented by two components.

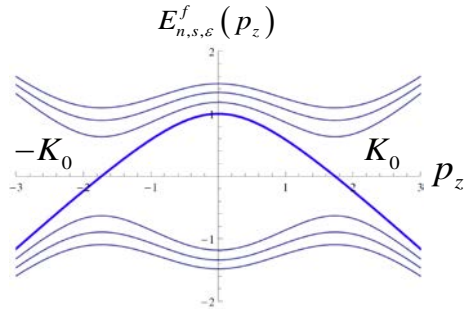

**Figure 4.** Energy spectrum as functions of $p_z$. Thick line denotes LLL.

One may immediately notice that the energy spectrum exhibits a peculiar feature: the spectrum for higher Landau levels with $n \neq 0$ is symmetric with respect to the level line, while that of LLL is asymmetric. Such spectral asymmetry of LLL gives rise to an interesting consequence. It has been shown in ref. [27] that such spectral asymmetry has a topological origin and is related to axial anomaly. Consequently it gives anomalous baryon number in the DCDW phase , and spreads the DCDW phase in the wide region in the QCD phase diagram in the presence of the magnetic field [24] . More interestingly, axial anomaly induces spontaneous magnetization in the DCDW phase [28], which in turn induces anomalous Hall current.

Writing the baryon number operator in the normal ordering form, $\widehat{N}_f = \frac{1}{2} \int d^3x [\Psi_f^\dagger(\mathbf{x}), \Psi_f(\mathbf{x})]$ ($f = u$ or $d$) we find $N(E) = N_{\text{norm}}(E) + N_{\text{anom}}$ by counting the number of the eigenstates below energy $E$. The first term is the usual one and can be written as $N_{\text{norm}}(E) = \int_0^E \text{Tr}\delta(\lambda - H)d\lambda$ at $T = 0$. The quasi-particle density of states can be written as

$$D^f_{\text{DCDW}}(\lambda) = N_c \frac{|e_f|B}{(2\pi)^2} \sum_{\varepsilon=\pm 1} \int_{-\infty}^{\infty} dp_z \left( \delta\left(\lambda - E_{n=0,\varepsilon}(p_z)\right) + \sum_{s=\pm 1} \sum_{n=1}^{\infty} \delta\left(\lambda - E^f_{n,s,\varepsilon}(p_z)\right) \right),$$ (27)

in the DCDW phase, and thereby $N_{\text{norm}}(E)$ reads

$$N_{\text{norm}}(E) = \sum_f N^f_{\text{norm}}(E) = \sum_f \int_0^E D^f_{\text{DCDW}}(\lambda)d\lambda.$$ (28)

Thus the contribution to the conductivity from the Fermi sea can be written as

$$\sigma_{xy}^{\text{Fermi}} = \sum_f e_f \left. \frac{\partial N^f_{\text{norm}}(E)}{\partial B} \right|_{E=\mu}.$$ (29)

The second one can be written as

$$N_{\text{anom}} = -\frac{1}{2}\int_{-\infty}^{+\infty} \text{sign}(\lambda)\text{Tr}\delta(\lambda - H)d\lambda, \tag{30}$$

and we can see that it comes from the spectral asymmetry; it can be also represented in terms of the $\eta$ invariant introduced by Atiyah-Patodi-Singer [29], $N_{\text{anom}} = -\frac{1}{2}\sum_f \eta_H^f$. The expression (30) is not well-defined as it is, and must be properly regularized to extract the physical result. Using Eq. (26) and the following gauge-invariant regularization, we can evaluate the $\eta$ invariant [27]

$$\eta_H^f = \lim_{s \to 0+} \frac{|e_f|B}{2\pi} N_c \int \frac{dp}{2\pi} \sum_\varepsilon \text{sign}(E_{n=0,\varepsilon}(p))|E_{n=0,\varepsilon}(p)|^{-s}$$

$$= \begin{cases} -N_c \dfrac{|e_f|Bq}{2\pi^2}, & M > \dfrac{q}{2} \\ -N_c \dfrac{|e_f|Bq}{2\pi^2} + \dfrac{|e_f|B}{\pi^2} N_c \sqrt{\left(\dfrac{q}{2}\right)^2 - M^2}, & M < \dfrac{q}{2} \end{cases}. \tag{31}$$

for each flavor. Accordingly, the anomalous Hall conductivity can be given as [30]

$$\sigma_{xy}^{\text{anom}} = -\frac{1}{2}\sum_f e_f \frac{\partial \eta_H^f}{\partial B}$$

$$= \begin{cases} \dfrac{e^2 q}{4\pi^2} & M > \dfrac{q}{2} \\ -\dfrac{e^2}{2\pi^2}\sqrt{\left(\dfrac{q}{2}\right)^2 - M^2} + \dfrac{e^2 q}{4\pi^2} & M < \dfrac{q}{2} \end{cases}. \tag{32}$$

This is the same result with the previous one, as it should be, but it is interesting to see the different manifestation of the topological effect in two calculations; one is given by the Berry curvature and Dirac monopole at the Weyl points, and the other by spectral asymmetry. It is to be noted again that our result is different from the one for WSM; some authors have obtained the different result [30]:

$$\sigma_{xy}^{\text{anom}} = -\frac{e^2}{2\pi^2}\sqrt{b^2 - m^2}, \tag{33}$$

for $b > m$ by using the momentum cut-off regularization for WSM. However, as is already mentioned, the $\eta$ invariant is given by the energy eigenstates, and gauge invariant regularizations must be used to obtain the relevant result. .

It may be worth noting that for quantized Hall effect in 2D Hall systems, $\sigma_{xy}^{\text{I}} = 0$, $\sigma_{xy}^{\text{II}} = -e^2 n/(2\pi)$ with *n* being the number of bands [26]. On the other hand, $\sigma_{xy}^{\text{II}} = 0$ in 3D Dirac materials with $b = 0$, and the whole contribution comes from $\sigma_{xy}^{\text{I}}, \sigma_{xy}^{\text{I}} \cong en_e/B$ with $n_e$ being the electron number [12], provided that the effect of impurities can be neglected.

## 5. Summary and Concluding remarks

We have discussed some topological aspects of the transport properties in the DCDW phase, in relation to WSM. Anomalous Hall conductivity is derived in two ways, using the Kubo formula. We have seen two different manifestation of the topological effect; one comes from the magnetic monopoles sitting at the Weyl points in the momentum space, and the other from the spectral asymmetry.

We have seen that the expression of the anomalous Hall conductivity is different from each other. We have not clearly seen its reason yet but probably the boundary effect must be taken into account to understand this difference.

Many subjects are left for further considerations. First of all, we must clarify the boundary effect for anomalous Hall effect. At the same time, we must clarify the role of axial anomaly in the presence of the magnetic field, since it should be closely related to the spectral asymmetry.

It is also interesting to extend the current framework so as to involve the CPT odd term, $\gamma_5 \boldsymbol{\gamma_5} \cdot \mathbf{b} \to \gamma_5 \gamma^\mu b_\mu$. Then we can incorporate the chiral magnetic effect [32], where $b_0$ plays a role of chiral chemical potential $\mu_5$.
It should be also interesting to study quantum phase transition between two cases, $b_0 > |\mathbf{b}|$ and $b_0 < |\mathbf{b}|$ indicated in [33].

As an application of our studies, one may consider the thermal evolution of compact stars. Since thermal conductivity is proportional to electric conductivity at low temperature by way of the Wiedemann-Franz law, we can discuss the effects of AHE during the thermal evolution of compact stars.

**Acknowledgments:** One of the authors (T.T.) thank the organizers for their hospitality during the conference. The work of H. A. was supported by JSPS KAKENHI Grant Number JP19K03868.